\documentclass[12pt]{article}
\usepackage{epsf}
\title{ {\bf Charged Lepton Flavor Physics and Extra Dimensions.}}
\author{\vspace{1cm}\\
        {\bf E. O. Iltan}
        \thanks{E-mail address:
        eiltan@newton.physics.metu.edu.tr}
 \\
        Physics Department, Middle East Technical University \\
        Ankara, Turkey\\}

\date{}

\begin{document}
\setlength{\baselineskip}{24pt}
\maketitle
\setlength{\baselineskip}{7mm}
\begin{abstract}
We estimate the charged lepton electric dipole moments and the
branching ratios of radiative lepton flavor violating decays in
the framework of the two Higgs doublet model with the inclusion
two extra dimensions. Here, we consider that the new Higgs doublet
is accessible to one of the extra dimensions with a Gaussian
profile and the fermions are accessible to the other extra
dimension with uniform zero mode profile. We observe that the
numerical values of the physical quantities studied enhance with
the additional effects due to the extra dimensions and they are
sensitive to the new Higgs localization.
\end{abstract}
\thispagestyle{empty}
\newpage
\setcounter{page}{1}
\section{Introduction}
The fermion electric dipole moments (EDMs) provide considerable
information about the CP violation since they are driven by the CP
violating interactions. In the standard model (SM), the  CP
violation is carried by the complex Cabibo Kobayashi Maskawa (CKM)
matrix elements in the quark sector. For the lepton sector, the
possible lepton mixing matrix is the candidate for non vanishing
CP violation. However, their negligibly small theoretical values
stimulate one to search new models beyond the SM and one can
obtain relatively greater EDMs with the extension of the particle
spectrum, such as multi Higgs doublet models (MHDM),
supersymmetric model (SUSY), \cite{Schmidt},..., etc.. Among
fermion EDMs, the charged lepton EDMs are worthwhile to study
since they are clean theoretically. In the literature, there exist
various experimental and theoretical work on the charged lepton
EDMs. The experimental results for electron, muon and tau read
$d_e =(1.8\pm 1.2\pm 1.0)\times 10^{-27} e\, cm$ \cite{Commins},
$d_{\mu} =(3.7\pm 3.4)\times 10^{-19} e\, cm$ \cite{Bailey} and
$d_{\tau}=3.1\times 10^{-16} e\, cm$ \cite{Groom}, respectively.
On the other hand, theoretically, the charged lepton EDMs have
been predicted extensively \cite {Bhaskar}-\cite
{IltanSplitHiggsLocalEDM}. They have been analyzed in the
framework of the seesaw model in \cite {Bhaskar}. In \cite
{Iltmuegam}, the EDMs of the charged leptons were estimated in the
2HDM and $d_e$ has been predicted at the order of the magnitude of
$10^{-32}\, e-cm$. The work \cite{IltanNonCom} was devoted to the
charged lepton EDMs in the framework of the SM with the inclusion
of non-commutative geometry and, in \cite{IltanExtrEDM}, the
effects of non-universal extra dimensions on these quantities in
the two Higgs doublet model were studied . Recently, the charged
lepton EDMs have been estimated in the split fermion scenario
\cite{IltanSplitEDM} and the effect of the localization of the new
Higgs doublet in the split fermion scenario has been analyzed in
\cite{IltanSplitHiggsLocalEDM}. In these theoretical works, some
of the possible models and the additional effects due to the
extension of the space-time have been studied to check the
possible enhancement in the numerical values of the EDMs. In
addition to the charged lepton EDMs the lepton flavor violating
(LFV) interactions are rich from the theoretical point of view
since they also exist at least at the loop level and make it
possible to search the free parameters of the models used. In the
SM, their branching ratios (BRs) are much below the experimental
limits since their existence depends on the neutrino mixing with
non zero neutrino masses and, therefore, it is worthwhile to study
them  in the models beyond the SM. The improvement of the
experimental measurements of the LFV processes make it possible to
understand the new physics effects more accurately. One of the
candidate model beyond the SM is the 2HDM, where the LFV
interactions are induced by the internal neutral Higgs bosons
$h^0$ and $A^0$. The $\mu\rightarrow e\gamma$ and $\tau\rightarrow
\mu\gamma$ are the examples of LFV interactions and the current
limits for their BRs are $1.2\times 10^{-11}$ \cite{Brooks} and
$3.9\times 10^{-7}$ \cite{Hayasaka}, respectively. A new
experiment at PSI has been described and aimed to reach to a
sensitivity of $BR\sim 10^{-14}$ for $\mu\rightarrow e\gamma$
decay \cite{Nicolo} and, at present, the experiment (PSI-R-99-05
Experiment) is still running in the MEG \cite{Yamada}. For
$\tau\rightarrow \mu\gamma$ decay an upper limit of $BR=9.0\,
(6.8)\, 10^{-8}$ at $90\%$ CL has been obtained \cite{Roney}
(\cite{Aubert}), which is an improvement almost by one order of
magnitude with respect to previous one. Besides the experimental
studies, there is an extensive work on the radiative LFV decays in
the literature from the theoretical point of view
\cite{Barbieri1}-\cite{Paradisi}. They are analyzed in the
supersymmetric models \cite{Barbieri1}-\cite{Barbieri1d},
\cite{Paradisi}, in a model independent way \cite{Chang}, in the
2HDM \cite{Chang, Iltan1, Iltan1a, Diaz, IltanExtrDim,
IltanLFVSplit, IltanLFVSplitFat, IltanLocalNewHiggs}.

This work is devoted to the prediction of the charged lepton EDMs
and the BRs of the LFV processes $\mu\rightarrow e\gamma$,
$\tau\rightarrow e\gamma$  and $\tau\rightarrow \mu\gamma$ in the
2HDM. In this model the CP violating nature of the charged lepton
EDMs are carried by the complex Yukawa couplings connected to the
new Higgs-lepton-lepton vertices and they are induced by the
internal new neutral Higgs bosons $h^0$ and $A^0$. Similarly, the
lepton flavor violation depends on these vertices and the
necessary loop diagrams contain internal neutral Higgs bosons
$h^0$ and $A^0$. Furthermore, we extend the space-time with the
additional two spatial dimensions and we consider that the new
Higgs doublet feels one of the extra dimensions (the sixth one)
with a Gaussian profile, and the fermions feel the other extra
dimension (the fifth one), with uniform zero mode profile.

The extra dimension scenario is a candidate as a possible solution
to the hierarchy problem of the SM. In the literature
\cite{Arkani}-\cite{Lam12}, the effects of extra dimensions on
various phenomena have been studied extensively. In the extra
dimension scenarios the compactification of extra dimension to a
circle $S^1$ with radius $R$ results in appearing new particles,
namely Kaluza-Klein (KK) modes in the theory. In the case that all
the fields feel the extra dimensions, so called universal extra
dimensions (UED), the extra dimensional momentum, therefore the KK
number at each vertex, is conserved. If the extra dimensions are
not felt by some fields in the theory, such type is called
non-universal extra dimensions where there is no restriction to
conserve the KK number at each vertex. This leads to the
possibility of the tree level interaction of KK modes with the
ordinary particles. In another scenario, so called the split
fermion scenario, the fermions are assumed to locate at different
points in the extra dimension with Gaussian profiles and the
hierarchy of fermion masses can be obtained from the overlaps of
fermion wave functions \cite{Mirabelli}-\cite{Grossman9}.

In the present work, we consider two extra dimensions and assume
that the new Higgs doublet is accessible to one of the extra
dimensions with a Gaussian profile around origin and also around
another point near to the origin. In addition to this, we take the
fermions are accessible to the other extra dimension with uniform
zero mode profile. Here the extra dimensions are compactified to
the orbifold $S^1/Z^2\times S^1$ so that the chiral structure of
four dimensional fermions are ensured. The localization of Higgs
bosons in the extra dimension has been considered previously.  The
idea of the localization of the SM Higgs, using the localizer
field, has been studied in \cite{Surujon}.
\cite{IltanSplitHiggsLocalEDM} was devoted to the localized new
Higgs scalars in the extra dimension where the localization is
measured by the strength of the small coupling of the localizer
field to the new Higgs scalar. In \cite{IltanLFVSplitFat}, the BRs
of the radiative LFV decays in the split fermion scenario, with
the assumption that the new Higgs doublet is restricted to the 4D
brane or to a part of the bulk in one and two extra dimensions, in
the framework of the 2HDM has been studied. In
\cite{IltanLocalNewHiggs,IltanLocalNewHiggsZl1l2} the new Higgs
doublet localization effects on radiative LFV decays and LFV Z
boson decays have been estimated.

Our analysis shows that the inclusion of lepton KK modes due to
the fifth dimension results in an enhancement in the charged
lepton EDMs and the BRs of LFV $l_i\rightarrow l_j \gamma$ decays.
Furthermore, we observe that these physical quantities are
strongly sensitive to the location of the new Higgs doublet
Gaussian profiles in the sixth dimension.

The paper is organized as follows: In Section 2, we present EDMs
of the charged leptons and the  BRs of the radiative LFV decays in
the 2HDM with the inclusion of two spatial extra dimensions.
Section 3 is devoted to discussion and our conclusions.
\section{ Charged Lepton Flavor Physics in the two Higgs doublet model
where the leptons and the new Higgs doublet feel different extra
dimensions.}
\subsection{Electric dipole moments of charged leptons}
The fermion EDM carries a valuable information about the existence
of the CP violation since  it emerges from the CP violating
fermion-fermion-photon interaction. The possible source of CP the
violation is the complex CKM matrix  (lepton mixing matrix)
elements for quarks (for leptons), in the framework of the SM.
However, their estimated numerical values are extremely small and
this makes it interesting to investigate new complex phases by
considering the physics beyond the SM. The extension of the Higgs
sector may bring additional complex phases with the assumption
that the flavor changing neutral currents (FCNC) are permitted at
tree level with new complex Yukawa couplings. The 2HDM is one of
the candidate to switch on the additional CP phase to enhance the
amount of the possible CP violation. On the other hand, the
addition of the spatial extra dimensions bring new contributions
which are sensitive to the compactification scale $1/R$ where $R$
is the radius of the compactification. In the present work, we
consider two additional dimensions and assume that the new Higgs
doublet feels one of the extra dimensions with a Gaussian profile,
and the fermions accessible to the other one.

The Yukawa Lagrangian responsible for the lepton EDM in a two
extra dimensions, respecting the considered scenario, reads:
\begin{eqnarray}
{\cal{L}}_{Y}=
\xi^{E}_{6 \,ij}\, \bar{l}_{i L}|_{z=0}\, \phi_{2}|_{y=0} \,E_{j
R}|_{z=0} + h.c. \,\,\, , \label{lagrangian}
\end{eqnarray}
where $L$ and $R$ denote chiral projections $L(R)=1/2(1\mp
\gamma_5)$. Here $l_{i L}$ ($E_{j R}$), with family indices $i,j$,
are the lepton doublets (singlets), $\phi_{2}$ is the new Higgs
doublet.We choose the Higgs doublets $\phi_{1}$ and $\phi_{2}$ as
\begin{eqnarray}
\phi_{1}=\frac{1}{\sqrt{2}}\left[\left(\begin{array}{c c}
0\\v+H^{0}\end{array}\right)\; + \left(\begin{array}{c c} \sqrt{2}
\chi^{+}\\ i \chi^{0}\end{array}\right) \right]\, ;
\phi_{2}=\frac{1}{\sqrt{2}}\left(\begin{array}{c c} \sqrt{2}
H^{+}\\ H_1+i H_2 \end{array}\right) \,\, . \label{choice}
\end{eqnarray}
with the vacuum expectation values,
\begin{eqnarray}
<\phi_{1}>=\frac{1}{\sqrt{2}}\left(\begin{array}{c c}
0\\v\end{array}\right) \,  \, ; <\phi_{2}>=0 \,\, ,
\label{choice2}
\end{eqnarray}
and collect SM (new) particles in the first (second) doublet.
Notice that $H_1$ and $H_2$ are the mass eigenstates $h^0$ and
$A^0$ respectively since no mixing occurs between two CP-even
neutral bosons $H^0$ and $h^0$ at tree level, in our case. Here,
we assume that the new Higgs scalars ($S=h^0, A^0$) are localized
in the extra dimension at the point $z_H$, $z_H=\alpha\, \sigma$
with Gaussian profiles,
\begin{eqnarray}
S(x,z)=A_H \,e^{-\beta (z-z_H)^2}\, S(x) \label{phi2H}\, ,
\end{eqnarray}
by an unknown mechanism\footnote{We consider the zero mode Higgs
scalars and we do not take into account the possible KK modes of
Higgs scalars since the mechanism for the localization is unknown
and we expect that the those contributions are small due to their
heavy masses.} with the normalization constant
\begin{eqnarray}
A_H=\frac{2\,( \beta)^{1/4}}{(2
\pi)^{1/4}\,\sqrt{Erf[\sqrt{2\,\beta}\,(\pi\,R+z_H)]+
Erf[\sqrt{2\,\beta}\,(\pi\,R-z_H)]}} \label{NormH} \, .
\end{eqnarray}
Here the parameter $\beta=1/\sigma^2$ regulates the amount of
localization, where $\sigma$, $\sigma=\rho\,R$, is the Gaussian
width of $S(x,z)$ in the extra dimension. The function $Erf[z]$ is
the error function, which is defined as
\begin{eqnarray}
Erf[z]=\frac{2}{\sqrt{\pi}}\,\int_{0}^{z}\,e^{-t^2}\,dt \,\, .
\label{erffunc}
\end{eqnarray}
On the other hand, the five dimensional lepton doublets and
singlets have both chiralities and the four dimensional Lagrangian
is constructed by expanding these  fields into their KK modes.
Besides, the fifth extra dimension denoted by $y$ is compactified
on an orbifold $S^1/Z_2$ with radius $R$. The KK decompositions of
the lepton  fields read
\begin{eqnarray}
l_i (x,y )& = & {1 \over {\sqrt{2 \pi R}}} \left\{ l_{i
L}^{(0)}(x) + \sqrt{2} \sum_{n=1}^{\infty} \left[l_{i L}^{(n)}(x)
 \cos(ny/R) + l_{i R}^{(n)}(x) \sin(ny/R)\right]\right\}\, ,\nonumber\\
E_{i}(x,y )& = & {1 \over {\sqrt{2 \pi R}}} \left\{ E_{i
R}^{(0)}(x) + \sqrt{2} \sum_{n=1}^{\infty}  \left[E_{i R}^{(n)}(x)
\cos(ny/R) + E_{i L}^{(n)}(x) \sin(ny/R)\right]\right\} \,\, ,
\label{f0}
\end{eqnarray}
where, $l_{i L}^{(0)}(x)$ and $E_{i R}^{(0)}(x)$ are the four
dimensional lepton doublets and lepton singlets respectively.

Now, we present the effective EDM interaction for a charged lepton
$l$ and it reads
\begin{eqnarray}
{\cal L}_{EDM}=i d_l \,\bar{l}\,\gamma_5 \,\sigma^{\mu\nu}\,l\,
F_{\mu\nu} \,\, , \label{EDM1}
\end{eqnarray}
where $F_{\mu\nu}$ is the electromagnetic field tensor, '$d_{l}$'
is EDM of the charged lepton $l$ and it is a real number by
hermiticity. In Fig. \ref{fig1} we present the 1-loop diagrams
which contribute to the EDMs of leptons with the help of the
complex Yukawa couplings. Here, we assume that there is no CKM
type lepton mixing matrix and, therefore, only the neutral Higgs
part gives a contribution to their EDMs. The complex Yukawa
couplings of the new Higgs doublet to the leptons  play the main
role in the determination of lepton EDM and they are modified with
the reduction of the extra dimensions. To obtain the
lepton-lepton-Higgs interaction coupling in four dimensions we
need to integrate the combination $\bar{l}^{(0
(n))}_{iL\,(R)}(x,y)\,S(x,z)\, l^{(n (0))}_{j R\,(l)}(x,y)$,
appearing in the part of the Lagrangian (eq. (\ref{lagrangian})),
over the fifth and sixth dimensions. Using the KK basis for lepton
fields (see eq. (\ref{f0})), we get
\begin{eqnarray}
\int_{-\pi R}^{\pi R}\, dz\,\,\int_{-\pi R}^{\pi R}\, dy\,\,
\delta(z)\, \delta(y)\, \bar{l}^{(0 (n))}_{iL\,(R)}(x,y)\,S(x,z)\,
l^{(n (0))}_{jR\,(L)}(x,y)=V_n \, \bar{l}^{(0(n))}_{iL\,(R)}(x)
\,S(x)\,\,l^{(n (0))}_{j R\,(L)}(x)\,\, , \label{intVij1}
\end{eqnarray}
where the factor $V_n$ reads
\begin{eqnarray}
V_n=\frac{A_H}{2\pi R}\, , \label{Vij1even}
\end{eqnarray}
and the function $A_H$ is defined in eq. (\ref{NormH}). Here, the
fields $l^{(n (0))}_{iL}$, $l^{(n (0))}_{i R}$ are four
dimensional left and right handed zero (n) mode lepton fields.
Here we define the Yukawa couplings in four dimensions as
\begin{eqnarray}
\xi^{E}_{ij}= V_n\, \xi^{E}_{6\, ij}\,\, , \label{coupl4}
\end{eqnarray}
where $\xi^{E}_{6\, ij}$ are  Yukawa couplings in six dimensions
(see eq. (\ref{lagrangian}))
\footnote{In the following we use the dimensionful complex
coupling $\bar{\xi}^{E}_{N}$ with the definition
$\xi^{E}_{N,ij}=\sqrt{\frac{4\, G_F}{\sqrt{2}}}\,
\bar{\xi}^{E}_{N,ij}$ where N denotes the word "neutral".} Notice
that we consider the compactification of two extra dimensions on
$(S^1/Z_2\times S^1)$.

Finally, the EDMs $d_l$ of charged leptons  $(l=e,\,\mu,\,\tau)$
can be calculated as a sum of contributions coming from neutral
Higgs bosons $h_0$ and $A_0$,
\begin{eqnarray}
d_l&=& -\frac{i G_F}{\sqrt{2}} \frac{e}{32\pi^2}\, Q_{\tau}\,c_H\,
((\bar{\xi}^{D\,*}_{N,l\tau})^2- (\bar{\xi}^{D}_{N,\tau
l})^2)\,\Bigg ( \frac{1}{m_{\tau}}\,
(F_1 (y_{h_0})-F_1 (y_{A_0}))\nonumber \\
&+& 2\, \sum_{n=1}^{\infty}\,
\frac{1}{\sqrt{m^2_{\tau}+m_n^2}}\,(F_1 (y_{n,h_0})-F_1
(y_{n,A_0})) \Bigg)\, , \label{emuEDM}
\end{eqnarray}
for $l=e,\mu$ and
\begin{eqnarray}
d_{\tau}&=& -\frac{i G_F}{\sqrt{2}} \frac{e}{32\pi^2}\,
Q_{\tau}\,c_H\, ((\bar{\xi}^{D\,*}_{N,\tau\tau})^2-
(\bar{\xi}^{D}_{N,\tau \tau})^2)\, \Bigg(\frac{1}{m_{\tau}}\, (F_2
(r_{h_0})-F_2(r_{A_0}))\nonumber \\ &+& 2\,
\sum_{n=1}^{\infty}\,\frac{1}{\sqrt{m^2_{\tau}+m_n^2}}\,(F_1
(y_{n,h_0})-F_1(y_{n,A_0}))\Bigg)
\,\, , \label{tauEDM}
\end{eqnarray}
where
\begin{eqnarray}
c_H =e^{\frac{-2\,z_H^2}{\sigma^2}}\, , \label{coeff}
\end{eqnarray}
for the case that the new Higgs scalars $S$ are localized around
the point $z_H$ different than origin. If the new Higgs
localization is around the origin $c_H$ reaches one. The functions
$F_1 (w)$, $F_2 (w)$ read
\begin{eqnarray}
F_1 (w)&=&\frac{w\,(3-4\,w+w^2+2\,ln\,w)}{(-1+w)^3}\nonumber \,\, , \\
F_2 (w)&=& w\, ln\,w + \frac{2\,(-2+w)\, w\,ln\,
\frac{1}{2}(\sqrt{w}-\sqrt{w-4})}{\sqrt{w\,(w-4)}}\,\, ,
\label{functions1}
\end{eqnarray}
with $y_{n,S}=\frac{m^2_{\tau}+m_n^2}{m^2_{S}}$,
$m_n=\frac{n}{R}$, $y_{S}=y_{0,S}$, $r_{S}=\frac{1}{y_{S}}$
and $Q_{\tau}$
is charge of $\tau$
lepton. In eq. (\ref{emuEDM}) and (\ref{tauEDM}) we take into
account only internal $\tau$-lepton contribution respecting our
assumption that the Yukawa couplings $\bar{\xi}^{E}_{N, ij},\,
i,j=e,\mu$, are small compared to $\bar{\xi}^{E}_{N,\tau\, i}\,
i=e,\mu,\tau$ due to the possible proportionality of the Yukawa
couplings to the masses of leptons in the vertices \cite{Sher}.
Here we used the parametrization
\begin{eqnarray}
\bar{\xi}^{E}_{N,\tau l}=|\bar{\xi}^{E}_{N,\tau l}|\, e^{i\,
\theta_{l}} \, . \label{xi2}
\end{eqnarray}
Therefore, the Yukawa factors in eqs. (\ref{emuEDM}),
(\ref{tauEDM}) can be written as
\begin{eqnarray}
((\bar{\xi}^{D\,*}_{N,l\tau})^2-(\bar{\xi}^{D}_{N,\tau
l})^2)=-2\,i \,sin\,2\theta_{l}\, |\bar{\xi}^{D}_{N,\tau l}|^2\, ,
\end{eqnarray}
where $l=e,\mu,\tau$ and $\theta_{l}$ is the CP violating
parameter which is the source of the lepton EDM. Notice that, we
make our calculations in arbitrary photon four momentum square
$q^2$ and take $q^2=0$ at the end.

\subsection{ The radiative LFV decays}
LFV $l_1\rightarrow l_2\gamma$ decays exist at loop level in the
SM and the numerical values of their BRs are far from the
experimental estimates. Therefore, one goes the models beyond
where the particle spectrum is extended and the additional
contributions result in an enhancement in the numerical values of
the physical parameters. Due to the extended Higgs sector, the
version of the 2HDM, permitting the existence of the FCNCs at tree
level, is one of the candidate to obtain relatively large BRs of
the decays under consideration. Furthermore, we take into account
the effects of two spatial extra dimensions which causes to
enhance the BRs due to the fact that the particle spectrum is
further extended after the compactification. Here, we consider the
effects of the additional Higgs sector with the assumption that
the new Higgs scalar zero modes are localized in one of the extra
dimension with Gaussian profiles by an unknown mechanism, on the
other hand, the zero modes of charged leptons have uniform profile
in the other extra dimension. The Yukawa Lagrangian responsible
for the LFV interactions in two extra dimensions are given in eq.
(\ref{lagrangian}).

\begin{figure}[htb]
\vskip -6.0truein \epsfxsize=4.8in\hskip 1.7truein
\leavevmode\epsffile{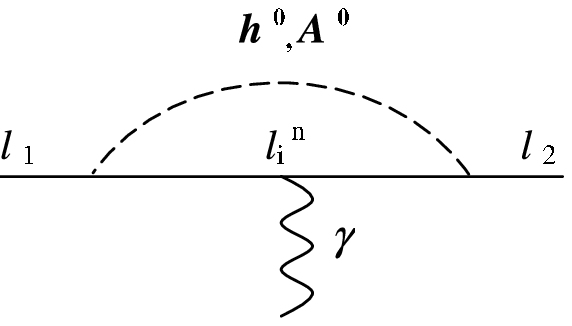} \vskip 10.0truein \caption[]{One
loop diagrams contribute to $l_1\rightarrow l_2 \gamma$ decay  due
to the zero mode neutral Higgs bosons $h^0$ and $A^0$ in the 2HDM,
for a single extra dimension. These diagrams contribute to EDM of
charged lepton $l_1$ for  $l_1=l_2$. Here $l_i^{(n)}$ represents
the internal KK mode charged lepton and n=0,1, ... Wavy lines
represent the electromagnetic field, dashed line the Higgs field,
solid line the charged leptons $l_{1 \,(i)}=e, \mu, \tau$.}
\label{fig1}
\end{figure}

Now, we will present the decay widths of the processes
$\mu\rightarrow e\gamma$, $\tau\rightarrow e\gamma$ and
$\tau\rightarrow \mu\gamma.$ Since they appear at least at one
loop level in the 2HDM (see Fig. \ref{fig1}) there exist the
logarithmic divergences in the calculations. These divergences can
be eliminated by using the on-shell renormalization
scheme\footnote{In this scheme, the self energy diagrams for
on-shell leptons vanish since they can be written as $
\sum(p)=(\hat{p}-m_{l_1})\bar{\sum}(p) (\hat{p}-m_{l_2})\, , $
however, the vertex diagrams (see Fig.\ref{fig1}) give non-zero
contribution. In this case, the divergences can be eliminated by
introducing a counter term $V^{C}_{\mu}$ with the relation
$V^{Ren}_{\mu}=V^{0}_{\mu}+V^{C}_{\mu} \, , $ where
$V^{Ren}_{\mu}$ ($V^{0}_{\mu}$) is the renormalized (bare) vertex
and by using the gauge invariance: $k^{\mu} V^{Ren}_{\mu}=0$.
Here, $k^\mu$ is the four momentum vector of the outgoing
photon.}. The decay width $\Gamma$ for the $l_1\rightarrow
l_2\gamma$ decay reads
\begin{eqnarray}
\Gamma (l_1\rightarrow l_2\gamma)=c_1(|A_1|^2+|A_2|^2)\,\, ,
\label{DWmuegam}
\end{eqnarray}
for $l_1\,(l_2)=\tau;\mu\,(\mu$ or $e; e)$. Here $c_1=\frac{G_F^2
\alpha_{em} m^3_{l_1}}{32 \pi^4}$, $A_1$ ($A_2$) is the left
(right) chiral amplitude and taking only $\tau$ lepton for the
internal line, they read
\begin{eqnarray}
A_1&=&Q_{\tau} \frac{1}{48\,m_{\tau}^2} \Bigg \{ 6\,m_\tau\,
\bar{\xi}^{E *}_{N,\tau l_2}\, \bar{\xi}^{E *}_{N,l_1\tau}\,c_H\,
\Bigg(  \Big( F_1 (y_{h^0})-F_1 (y_{A^0})\Big) \nonumber \\ &+&
2\, \sum_{n=1}^{\infty}\,
\frac{m_{\tau}}{\sqrt{m_{\tau}^2+m_n^2}}\,\Big( F_1 (y_{n,
h^0})-F_1 (y_{n, A^0})\Big ) \Bigg ) + m_{l_1}\,\bar{\xi}^{E
*}_{N,\tau l_2}\, \bar{\xi}^{E}_{N,\tau l_1}\,c_H\, \Bigg(
\Big( G (y_{h^0})+G(y_{A^0}) \Big )\nonumber \\
&+& 2\,\sum_{n=1}^{\infty}\, \frac{m^2_{\tau}}{m_{\tau}^2+m_n^2}\,
\Big( G (y_{n, h^0})+G (y_{n, A^0})\Big)\Bigg) \Bigg \}
\nonumber \,\, , \\
A_2&=&Q_{\tau} \frac{1}{48\,m_{\tau}^2} \Bigg \{ 6\,m_\tau\,
\bar{\xi}^{E}_{N, l_2 \tau}\, \bar{\xi}^{E}_{N,\tau l_1}\,c_H\,
\Bigg( \Big(F_1(y_{h^0})-F_1(y_{A^0})\Big)\nonumber \\
&+&
2\,\sum_{n=1}^{\infty}\,\frac{m_{\tau}}{\sqrt{m_{\tau}^2+m_n^2}}\,\Big(
F_1 (y_{n, h^0})-F_1 (y_{n, A^0})\Big) \Bigg)+
m_{l_1}\,\bar{\xi}^{E}_{N,l_2\tau}\, \bar{\xi}^{E *}_{N,l_1
\tau}\,c_H\, \Bigg(\Big( G (y_{h^0})+G (y_{A^0})\Big) \nonumber \\
&+&
2\,\sum_{n=1}^{\infty}\,\frac{m^2_{\tau}}{m_{\tau}^2+m_n^2}\,\Big(G
(y_{n, h^0})+ G (y_{n, A^0})\Big) \Bigg) \Bigg\}
 \,\, , \label{A1A22}
\end{eqnarray}
where $y_{n, S}=\frac{m^2_{\tau}+m_n^2}{m^2_{S}}$,
$m_n=\frac{n}{R}$ and $Q_{\tau}$ is the charge of $\tau$ lepton.
Here the vertex factor $c_H$  is defined in eq. (\ref{coeff}). The
function $F_1 (w)$ is given in eq. (\ref{functions1}) and $G (w)$
reads
\begin{eqnarray}
G (w)=\frac{w\,(2+3\,w-6\,w^2+w^3+ 6\,w\,ln\,w)}{(-1+w)^4} \,\, .
\label{functions2}
\end{eqnarray}
%
\section{Discussion}
In this work, we study  the EDMs of charged leptons and the BRs of
the LFV $l_1\rightarrow l_2\gamma$ decays in the 2HDM with the
addition of two spatial extra dimensions. Here we take that the
leptons feel the fifth dimension, and the new Higgs doublet is
localized with Gaussian profile in the sixth one. We consider that
the extra dimensions are compactified on $S_1/Z_2 \times S_1$ so
that the chiralities of four dimensional lepton fields are
guaranteed. On the other hand, we choose the location of the new
Higgs doublet around the origin and also at the point near to the
origin so that we estimate the possible effects coming from its
position in the extra dimension.

The existence of the lepton EDM interactions depend on the  CP
violating phases and, in the present work, we consider the complex
Yukawa couplings appearing in the FCNC at tree level in the
framework of the 2HDM. The leptonic complex Yukawa couplings
$\bar{\xi}^E_{N,ij}, i,j=e, \mu, \tau$ are set of free parameters
in the 2HDM and we consider the Yukawa couplings
$\bar{\xi}^{E}_{N,ij},\, i,j=e,\mu $, as smaller compared to
$\bar{\xi}^{E}_{N,\tau\, i}\, i=e,\mu,\tau$ and we assume that
$\bar{\xi}^{E}_{N,ij}$ is symmetric with respect to the indices
$i$ and $j$. Notice that the new Higgs masses are among the free
parameters and we take their numerical values as $m_{h^0}=100\,
GeV$, $m_{A^0}=200\, GeV$.

The inclusion of the spatial extra dimension that is felt by the
leptons brings new contributions due to their KK excitations. The
additional vertices appearing in the calculation of EDMs are
coming from the fermion-KK fermion-new Higgs interaction where the
KK number is not conserved. Here the compactification of fifth
dimension results in a new parameter, called the compactification
radius $R$, and it should take the numerical values not to
contradict with the experimental measurements. For the new Higgs
doublet, we consider the localization with Gaussian profile at any
point near to the origin in the sixth dimension. Therefore, the
localization width $\sigma=\rho\,R$ and the possible localization
point $z_H=\alpha\, \sigma$ is other free parameters which should
be examined. The direct limits from searching for KK gauge bosons
imply $1/R> 800\,\, GeV$, the precision electro weak bounds on
higher dimensional operators generated by KK exchange place a far
more stringent limit $1/R> 3.0\,\, TeV$ \cite{Rizzo} and, from
$B\rightarrow \phi \, K_S$, the lower bounds for the scale $1/R$
have been obtained as $1/R > 1.0 \,\, TeV$, from $B\rightarrow
\psi \, K_S$ one got $1/R
> 500\,\, GeV$, and from the upper limit of the $BR$, $BR \, (B_s
\rightarrow \mu^+ \mu^-)< 2.6\,\times 10^{-6}$, the estimated
limit was $1/R > 800\,\, GeV$ \cite{Hewett}. Here we take the
compactification scale in the broad range, $100 \,GeV \geq 1/R
\geq 1000 \,GeV$, the width $\sigma=\rho\,R$ with $\rho \sim
0.001$ and the parameter $\alpha$, which regulates the
localization point $z_H=\alpha\,\sigma$, in the interval $0.001
\geq \alpha \geq 1$.

Now, we start to estimate the charged lepton EDMs and to study the
compactification scale $1/R$ and the new Higgs location point
dependencies of these measurable quantities.

\begin{figure}[htb]
\vskip -3.0truein \centering \epsfxsize=6.8in
\leavevmode\epsffile{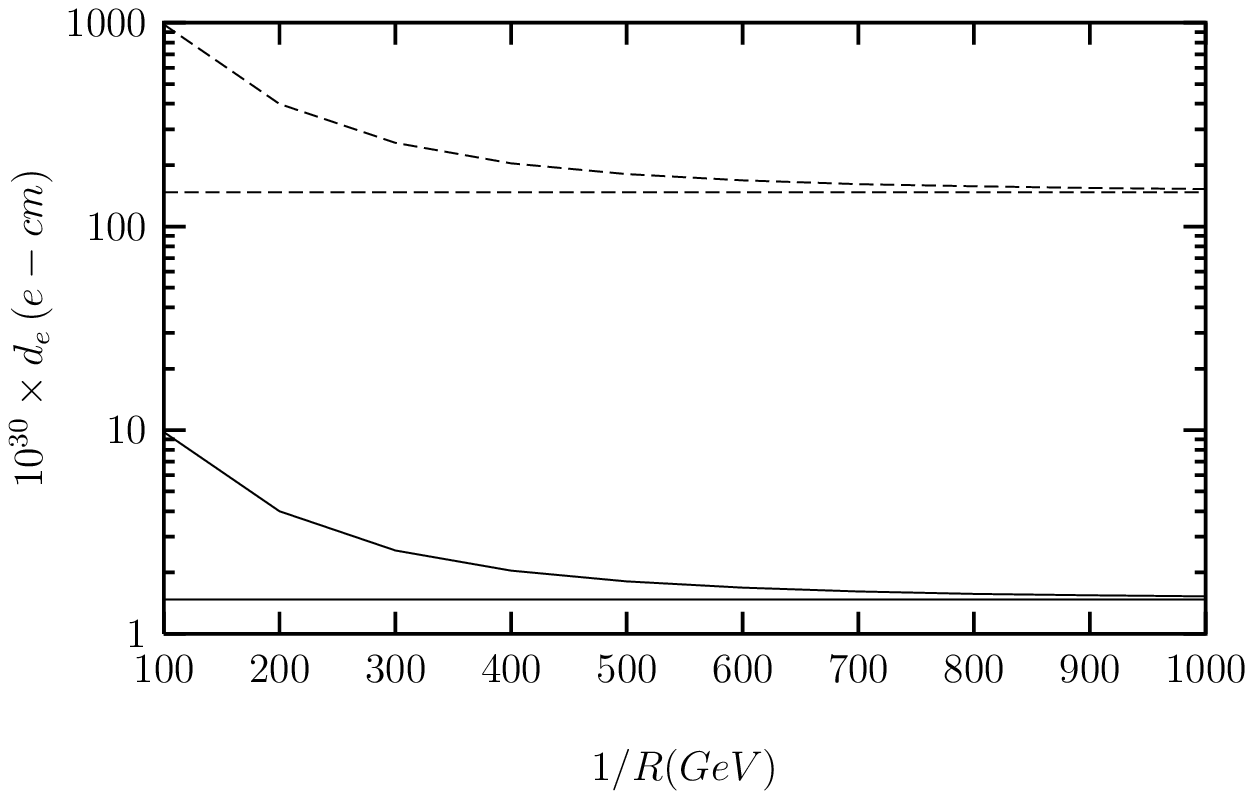} \vskip -3.0truein
\caption[]{$d_{e}$ with respect to $1/R$ and $sin\,\theta_e=0.5$.
Here the solid-dashed line (curve) represents the EDM for
$\bar{\xi}^{E}_{N,\tau e} =0.001-0.01\, GeV$ without (with) lepton
KK mode contribution in the case that the new Higgs doublet is
located around the origin in the sixth dimension.} \label{EDMeR}
\end{figure}

\begin{figure}[htb]
\vskip -3.0truein \centering \epsfxsize=6.8in
\leavevmode\epsffile{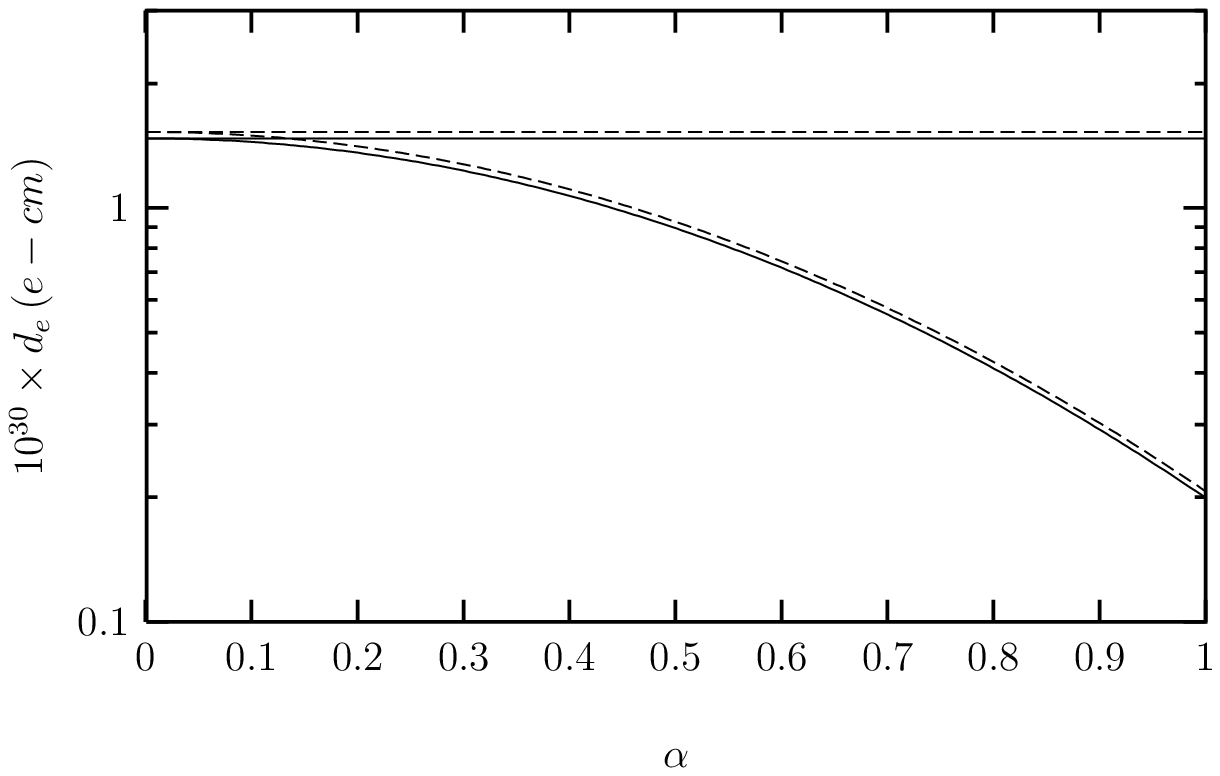} \vskip -3.0truein
\caption[]{$d_e$ with respect to $\alpha$  for $1/R= 1000\,GeV$
and $\bar{\xi}^{E}_{N,\tau e} =0.001\, GeV$. Here the solid-dashed
line (curve) represents the EDM without-with lepton KK mode
contribution in the case that the new Higgs scalars are located
around the origin ($z_H=\alpha\,\sigma$) in the sixth dimension.}
\label{EDMealf}
\end{figure}

In  Fig. \ref{EDMeR}, we plot EDM $d_e$ with respect to the scale
$1/R$  for the intermediate value $sin\,\theta_e=0.5$. Here the
solid-dashed line (curve) represents the EDM for
$\bar{\xi}^{E}_{N,\tau e} =0.001-0.01\, GeV$ without (with) lepton
KK mode contribution in the case that the new Higgs doublet is
located around the origin in the sixth dimension. The electron EDM
is at the order of magnitude of $10^{-28}\, (e-cm)$ for the
coupling $\bar{\xi}^{E}_{N,\tau e} =0.01\, GeV$ and enhances at
the order of $25\%$ for the compactification scale $1/R\sim
500\,GeV$. We study the effect of the different location of the
Gaussian profile of the new Higgs doublet on $d_e$ by plotting
this quantity with respect to the $\alpha$ for $1/R= 1000\,GeV$
and $\bar{\xi}^{E}_{N,\tau e} =0.001\, GeV$ (see Fig.
\ref{EDMealf}). Here the solid-dashed line (curve) represents the
EDM without-with lepton KK mode contribution in the case that the
new Higgs scalars are located around the origin
($z_H=\alpha\,\sigma$) in the sixth dimension. We observe that the
EDM is suppressed almost one order of magnitude even in the case
that the Gaussian profiles are located one $\sigma$ farther from
the origin. This shows that the EDM is strongly sensitive to the
location of the Gaussian profiles of new Higgs scalars in the
sixth dimension.

\begin{figure}[htb]
\vskip -3.0truein \centering \epsfxsize=6.8in
\leavevmode\epsffile{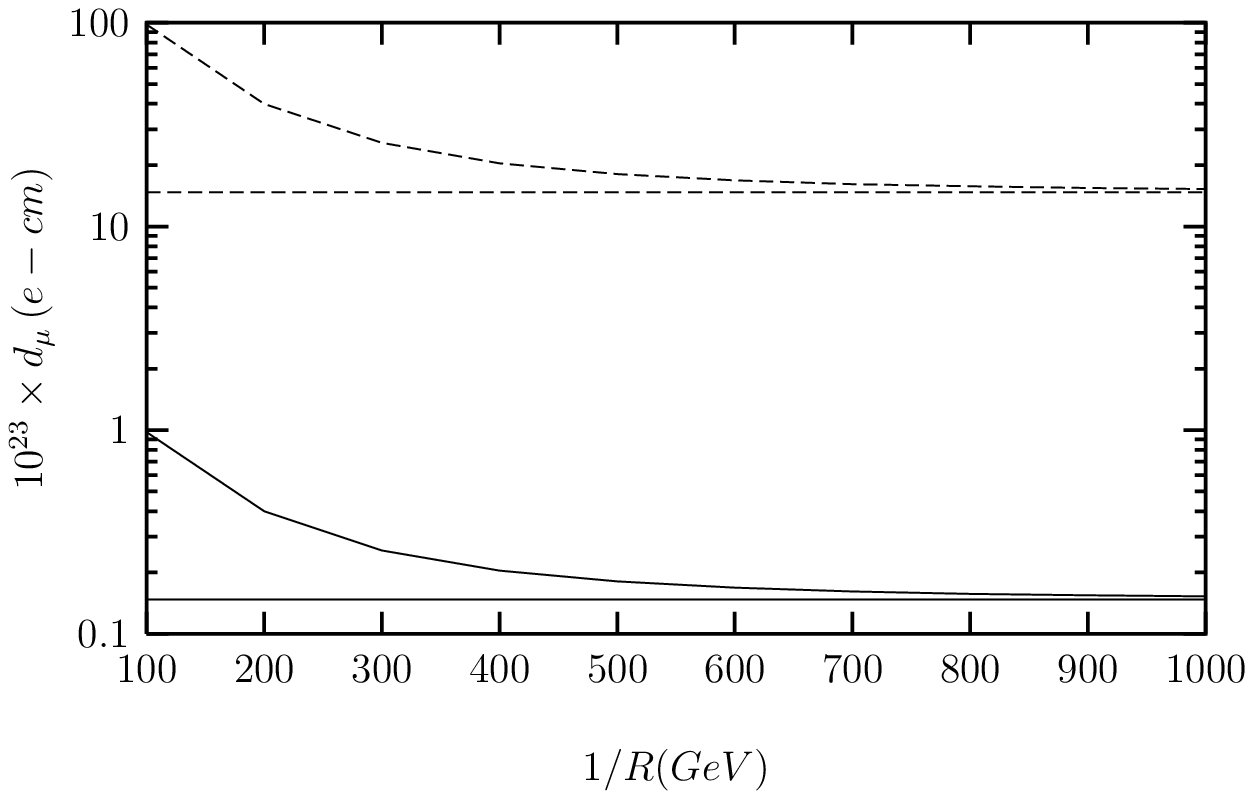} \vskip -3.0truein
\caption[]{$d_{\mu}$ with respect to $1/R$ for
$sin\,\theta_\mu=0.5$. Here the solid-dashed line (curve)
represents the EDM for $\bar{\xi}^{E}_{N,\tau \mu} =1-10\, GeV$
without (with) lepton KK mode contribution in the case that the
new Higgs doublet is located around the origin in the sixth
dimension.} \label{EDMmuR}
\end{figure}

\begin{figure}[htb]
\vskip -3.0truein \centering \epsfxsize=6.8in
\leavevmode\epsffile{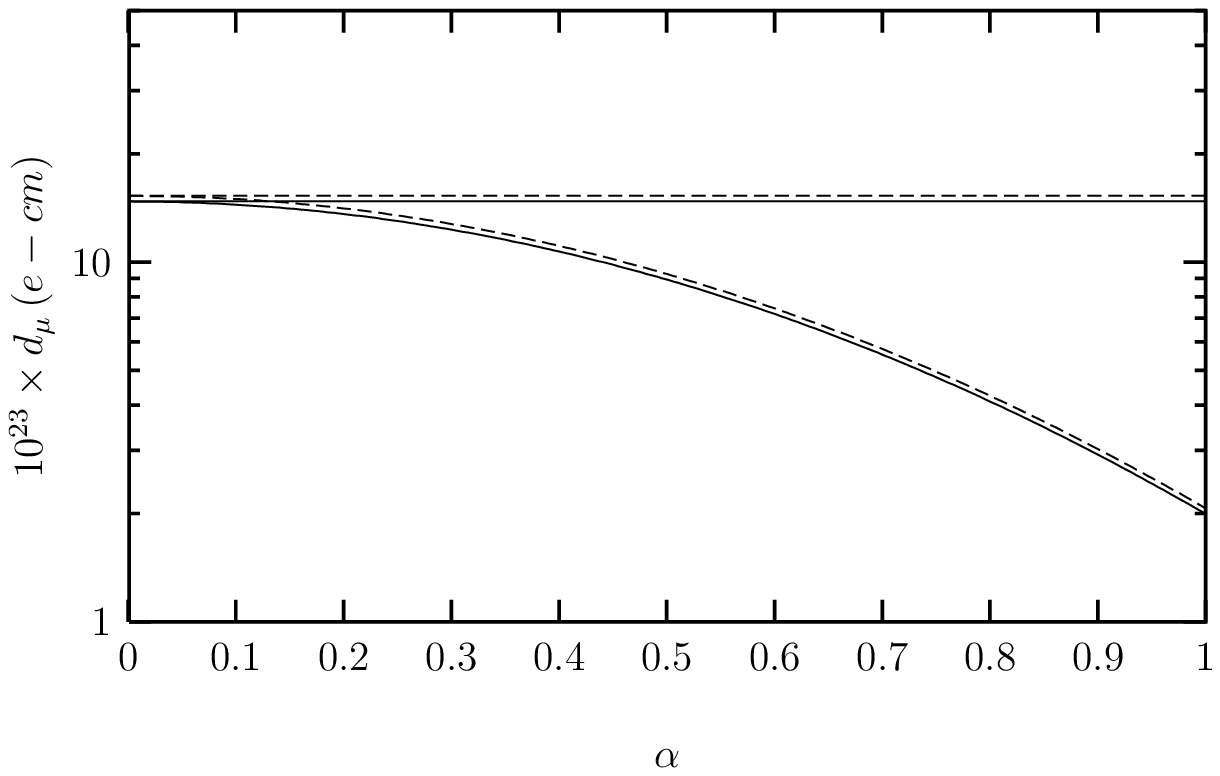} \vskip -3.0truein
\caption[]{$d_\mu$ with respect to $\alpha$  for $1/R= 1000\,GeV$
and $\bar{\xi}^{E}_{N,\tau \mu} =10\, GeV$. Here the solid-dashed
line (curve) represents the EDM without-with lepton KK mode
contribution in the case that the new Higgs scalars are located
around the origin ($z_H=\alpha\,\sigma$) in the sixth dimension.}
\label{EDMmualf}
\end{figure}

Fig. \ref{EDMmuR} is devoted to EDM $d_\mu$ with respect to the
scale $1/R$ for the intermediate value $sin\,\theta_\mu=0.5$. Here
the solid-dashed line (curve) represents the EDM for
$\bar{\xi}^{E}_{N,\tau \mu} =1-10\, GeV$ without (with) lepton KK
mode contribution in the case that the new Higgs doublet is
located around the origin in the sixth dimension. The muon EDM is
at the order of magnitude of $10^{-22}\, (e-cm)$ for the coupling
$\bar{\xi}^{E}_{N,\tau \mu} =10\, GeV$ and enhances nearly $30\%$
for the compactification scale $1/R\sim 500\,GeV$. For the small
values of the compactification scale the enhancement is almost one
order of magnitude. We represent the effect of the different
location of the Gaussian profile of the new Higgs doublet on
$d_\mu$ in Fig. \ref{EDMmualf}, by taking $1/R= 1000\,GeV$ and
$\bar{\xi}^{E}_{N,\tau \mu} =10\, GeV$. Here the solid-dashed line
(curve) represents the EDM without-with lepton KK mode
contribution in the case that the new Higgs scalars are located
around the origin ($z_H=\alpha\,\sigma$) in the sixth  dimension.
We observe that the EDM is strongly sensitive to the location of
the Gaussian profiles of new Higgs scalars in the sixth dimension
and its magnitude decreases almost one order of magnitude for the
case that the Gaussian profiles are located at most $\sigma$
farther from the origin.

\begin{figure}[htb]
\vskip -3.0truein \centering \epsfxsize=6.8in
\leavevmode\epsffile{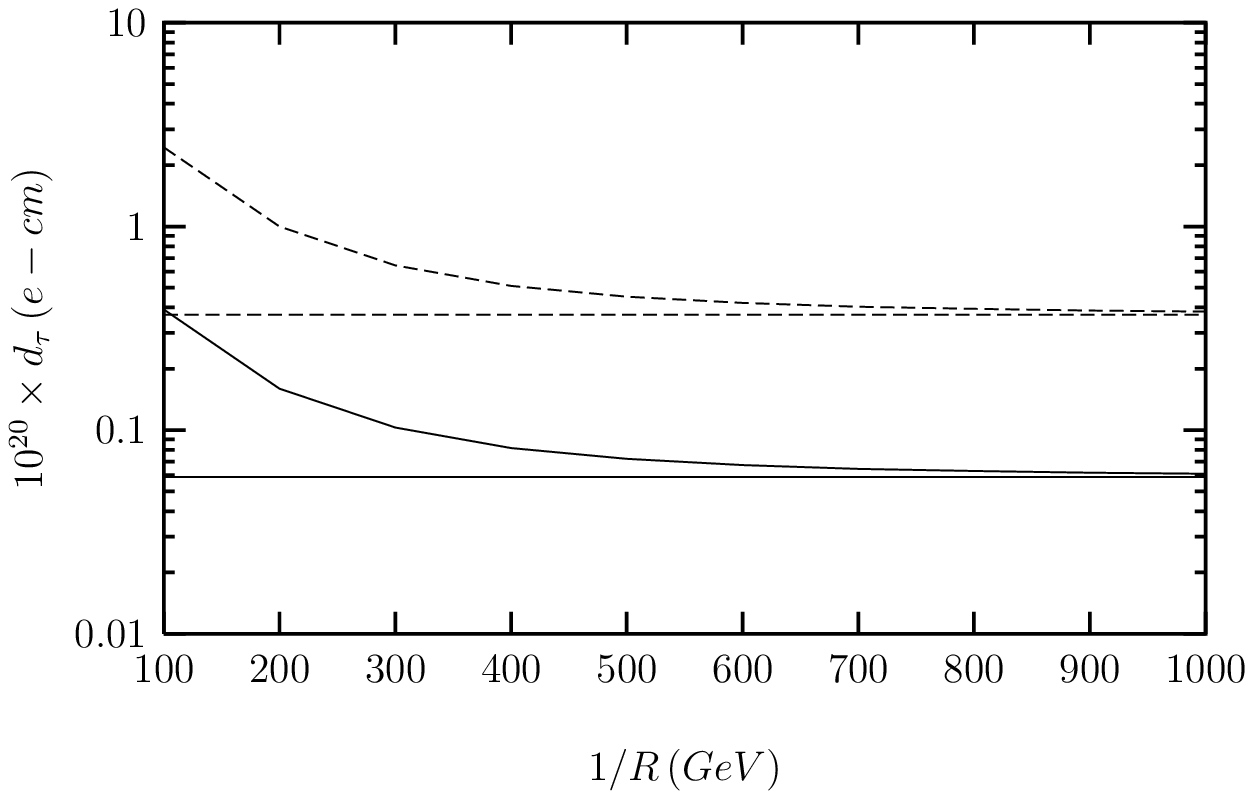} \vskip -3.0truein
\caption[]{$d_{\tau}$ with respect to $1/R$ for
$sin\,\theta_\tau=0.5$. Here the solid-dashed line (curve)
represents the EDM for $\bar{\xi}^{E}_{N,\tau \tau} =20-50\, GeV$
without (with) lepton KK mode contribution in the case that the
new Higgs doublet is located around the origin in the sixth
dimension.} \label{EDMtauR}
\end{figure}

\begin{figure}[htb]
\vskip -3.0truein \centering \epsfxsize=6.8in
\leavevmode\epsffile{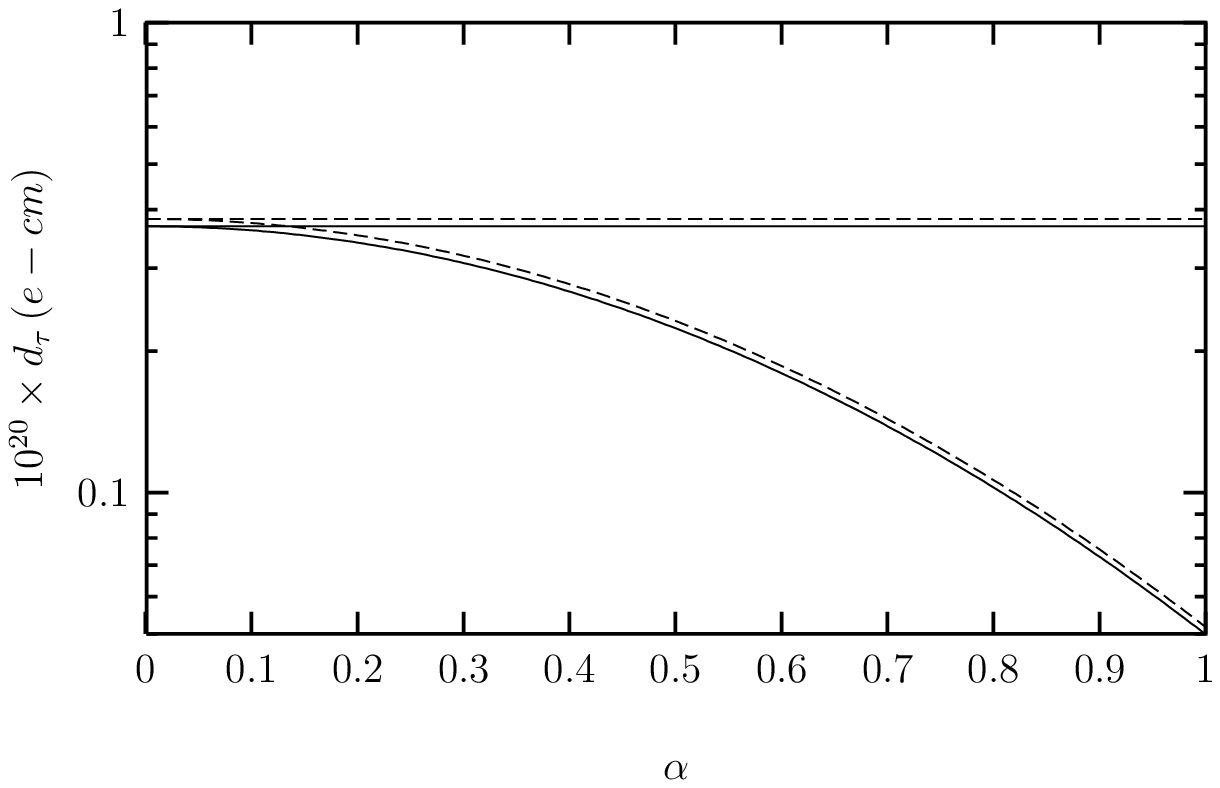} \vskip -3.0truein
\caption[]{$d_\tau$ with respect to $\alpha$  for $1/R= 1000\,GeV$
and $\bar{\xi}^{E}_{N,\tau \tau} =50\, GeV$. Here the solid-dashed
line (curve) represents the EDM without-with lepton KK mode
contribution in the case that the new Higgs scalars are located
around the origin ($z_H=\alpha\,\sigma$) in the sixth dimension.}
\label{EDMtaualf}
\end{figure}

Finally, we make the same analysis for $\tau$ lepton EDM $d_\tau$.
In Fig. \ref{EDMtauR} we present $d_\tau$ with respect to the
scale $1/R$ for the intermediate value $sin\,\theta_\mu=0.5$. Here
the solid-dashed line (curve) represents the EDM for
$\bar{\xi}^{E}_{N,\tau \tau} =20-50\, GeV$ without (with) lepton
KK mode contribution in the case that the new Higgs doublet is
located around the origin in the sixth dimension. $d_\tau$ is at
the order of magnitude of $10^{-20}\, (e-cm)$ for the coupling
$\bar{\xi}^{E}_{N,\tau \tau} =50\, GeV$ and enhances more than
$20\%$ for the compactification scale $1/R\sim 500\,GeV$. For the
small values of the compactification scale the enhancement is
almost one order of magnitude.  The effect of the different
locations of the Gaussian profile of the new Higgs doublet on
$d_\tau$ is represented in Fig. \ref{EDMtaualf}. In this figure we
take $1/R= 1000\,GeV$ and $\bar{\xi}^{E}_{N,\tau \tau} =50\, GeV$.
Here the solid-dashed line (curve) represents the EDM without-with
lepton KK mode contribution in the case that the new Higgs doublet
is located around the origin ($z_H=\alpha\,\sigma$) in the sixth
dimension. We observe that the EDM is strongly sensitive to the
location of new Higgs doublet Gaussian profiles in the sixth
dimension and its magnitude is suppressed up to values less than
one order of magnitude for the case that the Gaussian profile is
located at most one $\sigma$ farther from the origin.

Now we continue our analysis for the BRs of the LFV decays
$l_1\rightarrow l_2 \gamma$. First, we consider that the new Higgs
is localized around the origin in the extra dimension.
Furthermore, we choose the localization point is near to the
origin and study its effect on the BRs.

\begin{figure}[htb]
\vskip -3.0truein \centering \epsfxsize=6.8in
\leavevmode\epsffile{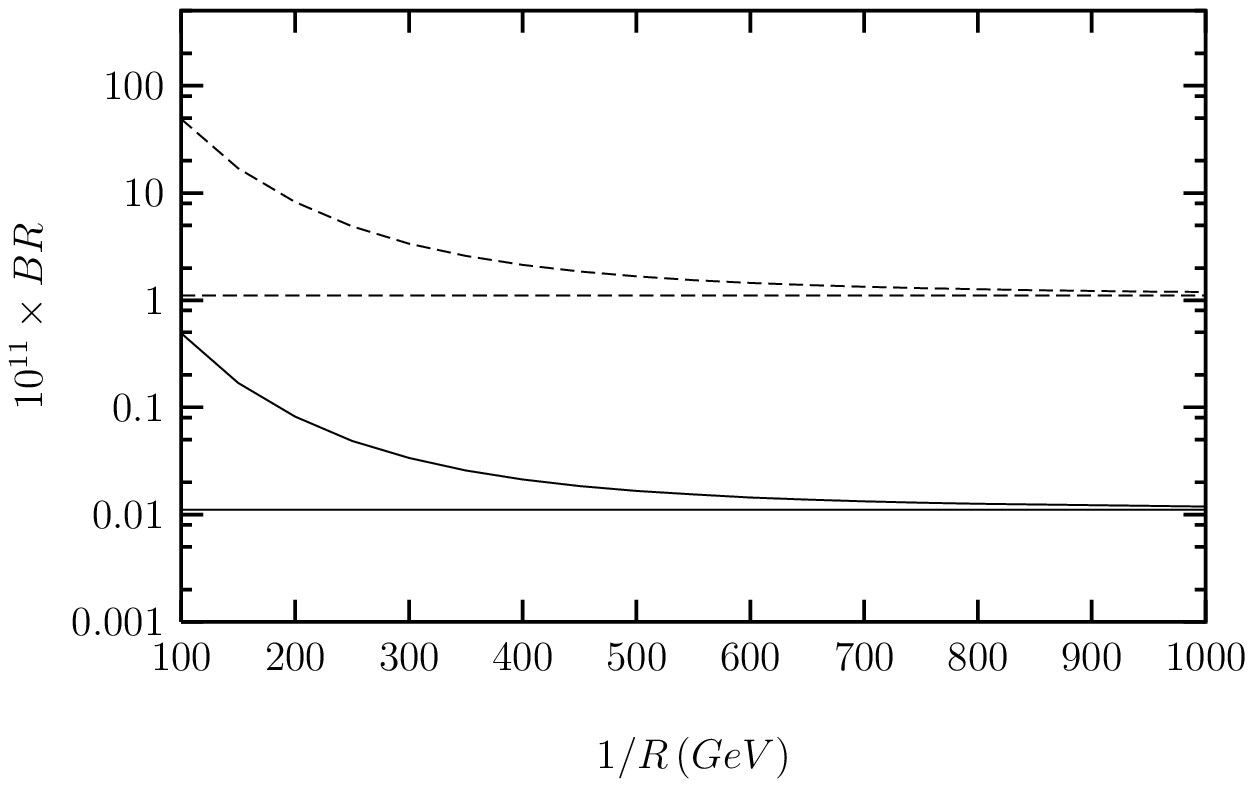} \vskip -3.0truein
\caption[]{BR($\mu\rightarrow e \gamma$) with respect to $1/R$.
Here the solid-dashed line (curve) represents the BR for
$\bar{\xi}^{E}_{N,\tau e} =0.001\, GeV$ and $\bar{\xi}^{E}_{N,\tau
\mu} =1\, GeV$-$\bar{\xi}^{E}_{N,\tau \mu} =10\, GeV$ without
(with) lepton KK mode contribution in the case that the new Higgs
doublet is located around the origin in the sixth dimension.}
\label{BRmuegamR}
\end{figure}

\begin{figure}[htb]
\vskip -3.0truein \centering \epsfxsize=6.8in
\leavevmode\epsffile{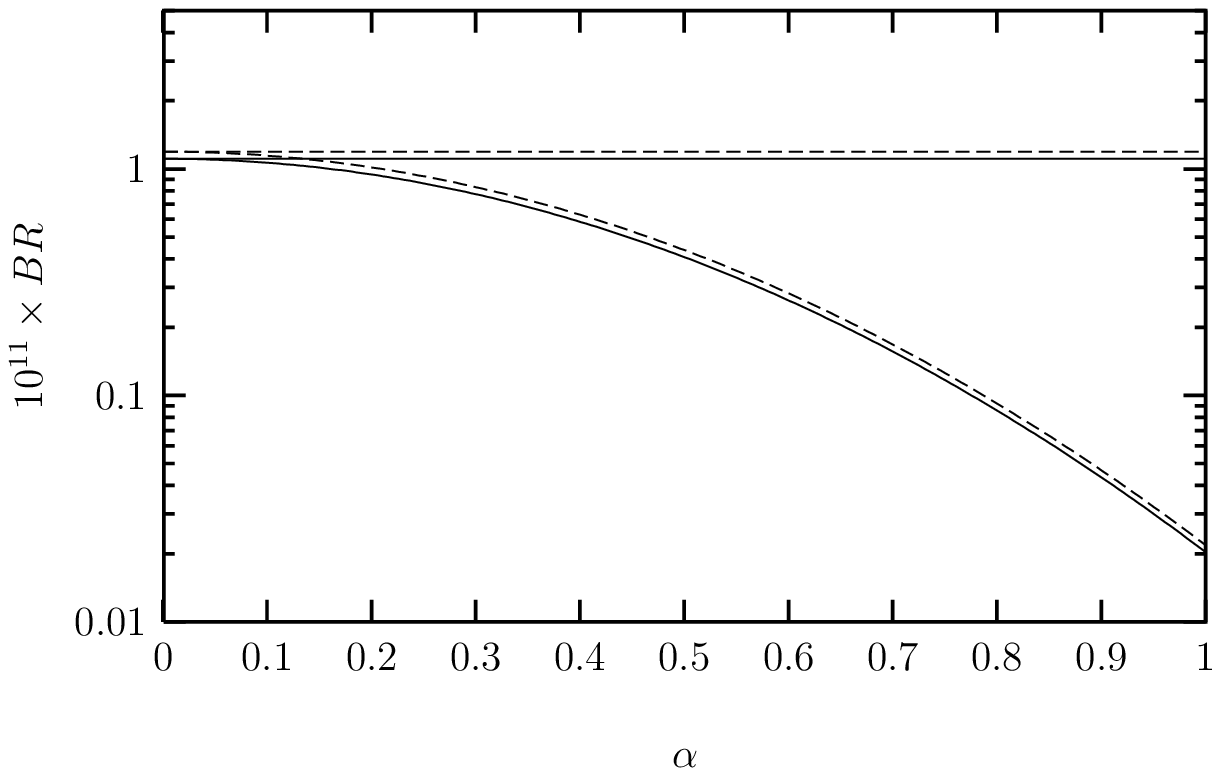} \vskip -3.0truein
\caption[]{BR($\mu\rightarrow e \gamma$) with respect to $\alpha$
for $1/R= 1000\,GeV$ and $\bar{\xi}^{E}_{N,\tau e} =0.001\, GeV$,
$\bar{\xi}^{E}_{N,\tau \mu} =10\, GeV$. Here the solid-dashed line
(curve) represents the BR without-with lepton KK mode contribution
in the case that the new Higgs doublet is located around the
origin ($z_H=\alpha\,\sigma$) in the sixth dimension.}
\label{BRmuegamalf}
\end{figure}

In  Fig. \ref{BRmuegamR}, we plot BR ($\mu\rightarrow e \gamma$)
with respect to the scale $1/R$. Here the solid-dashed line
(curve) represents the BR for $\bar{\xi}^{E}_{N,\tau e} =0.001\,
GeV$ and $\bar{\xi}^{E}_{N,\tau \mu} =1\,
GeV$-$\bar{\xi}^{E}_{N,\tau \mu} =10\, GeV$ without (with) lepton
KK mode contribution in the case that the new Higgs doublet is
located around the origin in the sixth dimension. The
BR($\mu\rightarrow e \gamma$) is at the order of magnitude of
$10^{-11}$ for the coupling $\bar{\xi}^{E}_{N,\tau \mu} =10\,
GeV$. The inclusion of internal lepton KK modes results in $50\%$
enhancement in the BR for the compactification scale $1/R\sim
500\,GeV$. This enhancement is more than one order of magnitude
for the small values of the scale $1/R$. Fig. \ref{BRmuegamalf} is
devoted to the new Higgs doublet Gaussian profile location scale
$\alpha$ dependence of the BR ($\mu\rightarrow e \gamma$) for
$1/R= 1000\,GeV$ and $\bar{\xi}^{E}_{N,\tau e} =0.001\, GeV$,
$\bar{\xi}^{E}_{N,\tau \mu} =10\, GeV$. Here the solid-dashed line
(curve) represents the BR without-with lepton KK mode contribution
in the case that the new Higgs is located around the origin
($z_H=\alpha\,\sigma$) in the sixth dimension. We observe that the
BR is strongly sensitive to the location of new Higgs scalar
Gaussian profiles in the sixth dimension and it is suppressed
almost two orders of magnitude even in the case that the Gaussian
profile is located one $\sigma$ farther from the origin.

\begin{figure}[htb]
\vskip -3.0truein \centering \epsfxsize=6.8in
\leavevmode\epsffile{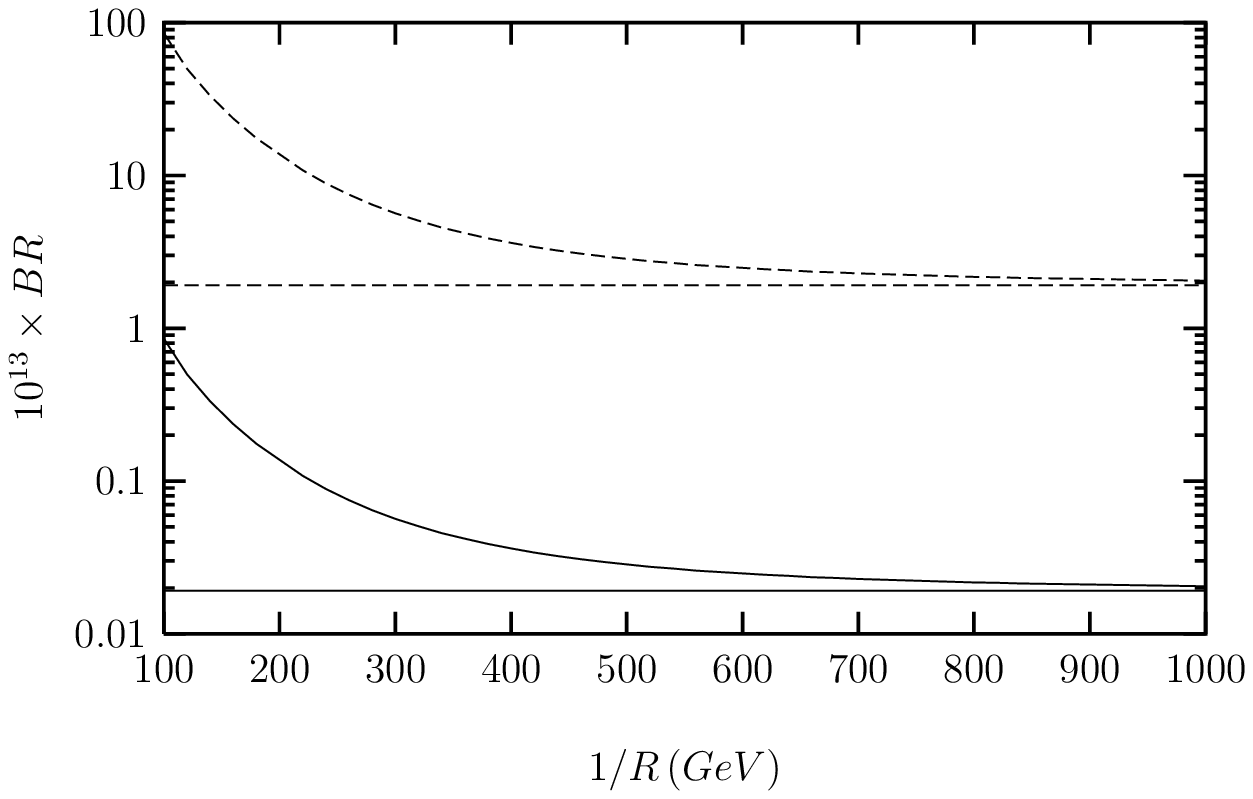} \vskip -3.0truein
\caption[]{BR($\tau\rightarrow e \gamma$) with respect to $1/R$.
Here the solid-dashed line (curve) represents the BR for
$\bar{\xi}^{E}_{N,\tau \tau} =50\, GeV$ and $\bar{\xi}^{E}_{N,\tau
e} =0.001\, GeV$ -$\bar{\xi}^{E}_{N,\tau e} =0.01\, GeV$  without
(with) lepton KK mode contribution in the case that the new Higgs
doublet is located around the origin in the sixth dimension.}
\label{BRtauegamR}
\end{figure}

\begin{figure}[htb]
\vskip -3.0truein \centering \epsfxsize=6.8in
\leavevmode\epsffile{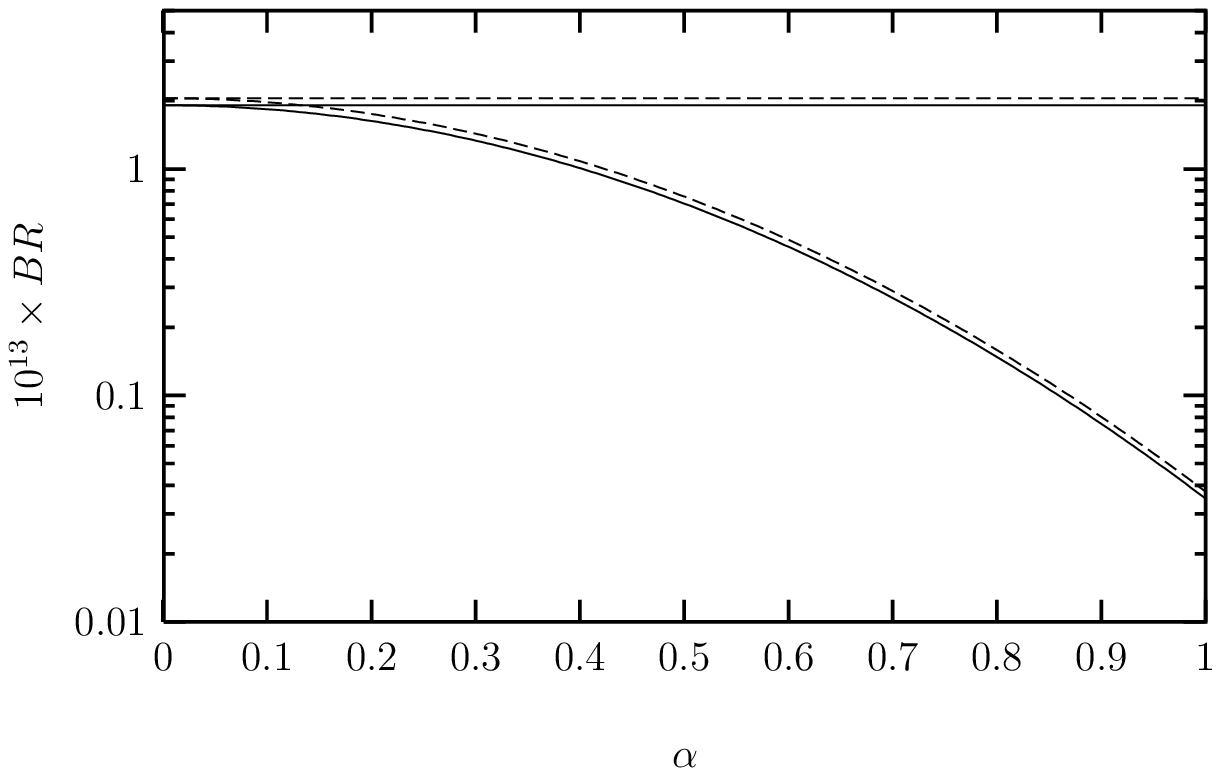} \vskip -3.0truein
\caption[]{BR($\tau\rightarrow e \gamma$) with respect to $\alpha$
for $1/R= 1000\,GeV$ and $\bar{\xi}^{E}_{N,\tau e} =0.01\, GeV$,
$\bar{\xi}^{E}_{N,\tau \tau} =50\, GeV$. Here the solid-dashed
line (curve) represents the BR without-with lepton KK mode
contribution in the case that the new Higgs doublet is located
around the origin ($z_H=\alpha\,\sigma$) in the sixth dimension.}
\label{BRtauegamalf}
\end{figure}

Fig. \ref{BRtauegamR} is devoted to the BR ($\tau\rightarrow e
\gamma$) with respect to the scale $1/R$. Here the solid-dashed
line (curve) represents the BR for $\bar{\xi}^{E}_{N,\tau \tau}
=50\, GeV$ and $\bar{\xi}^{E}_{N,\tau e} =0.001\, GeV$
-$\bar{\xi}^{E}_{N,\tau e} =0.01\, GeV$ without (with) lepton KK
mode contribution in the case that the new Higgs is located around
the origin in the sixth dimension. The BR ($\tau\rightarrow e
\gamma$) is at the order of magnitude of $10^{-13}$ for the
coupling $\bar{\xi}^{E}_{N,\tau e} =0.01\, GeV$. The additional
contribution coming from the internal lepton KK modes causes
$50\%$ enhancement in the BR for the compactification scale
$1/R\sim 500\,GeV$ and the BR increases up to the values
$10^{-11}$ for the small values of the scale $1/R$. Fig.
\ref{BRtauegamalf} represents the scale $\alpha$ dependence of the
BR($\tau\rightarrow e \gamma$) for $1/R= 1000\,GeV$ and
$\bar{\xi}^{E}_{N,\tau e} =0.01\, GeV$, $\bar{\xi}^{E}_{N,\tau
\tau} =50\, GeV$. Here the solid-dashed line (curve) represents
the BR without-with lepton KK mode contribution in the case that
the new Higgs is located around the origin ($z_H=\alpha\,\sigma$)
in the sixth dimension. We observe that the BR  is suppressed
almost two orders of magnitude for the case that the Gaussian
profile is located one $\sigma$ farther from the origin, similar
to the $\mu\rightarrow e \gamma$ decay.

\begin{figure}[htb]
\vskip -3.0truein \centering \epsfxsize=6.8in
\leavevmode\epsffile{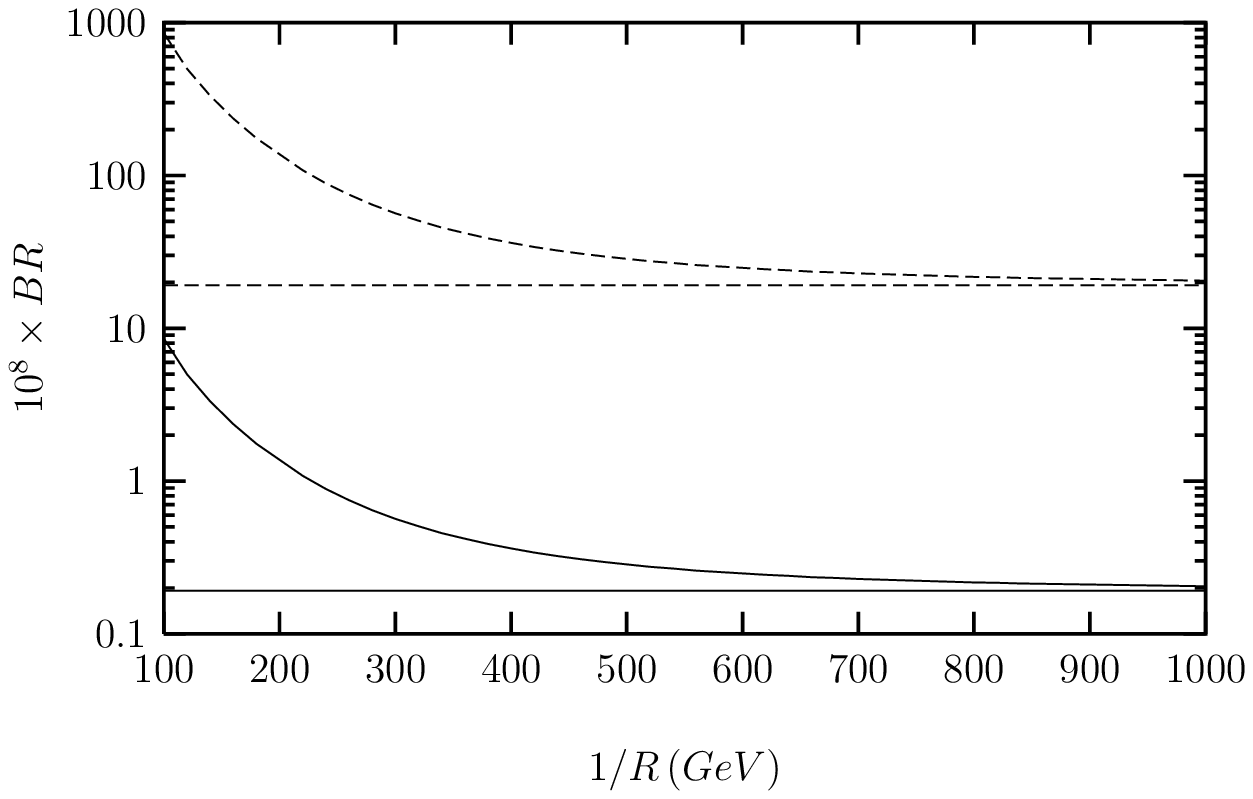} \vskip -3.0truein
\caption[]{BR($\tau\rightarrow \mu \gamma$) with respect to $1/R$.
Here the solid-dashed line (curve) represents the BR for
$\bar{\xi}^{E}_{N,\tau \tau} =50\, GeV$ and $\bar{\xi}^{E}_{N,\tau
\mu} =1\, GeV$ -$\bar{\xi}^{E}_{N,\tau \mu} =10\, GeV$ without
(with) lepton KK mode contribution in the case that the new Higgs
doublet is located around the origin in the sixth dimension.}
\label{BRtaumugamR}
\end{figure}

\begin{figure}[htb]
\vskip -3.0truein \centering \epsfxsize=6.8in
\leavevmode\epsffile{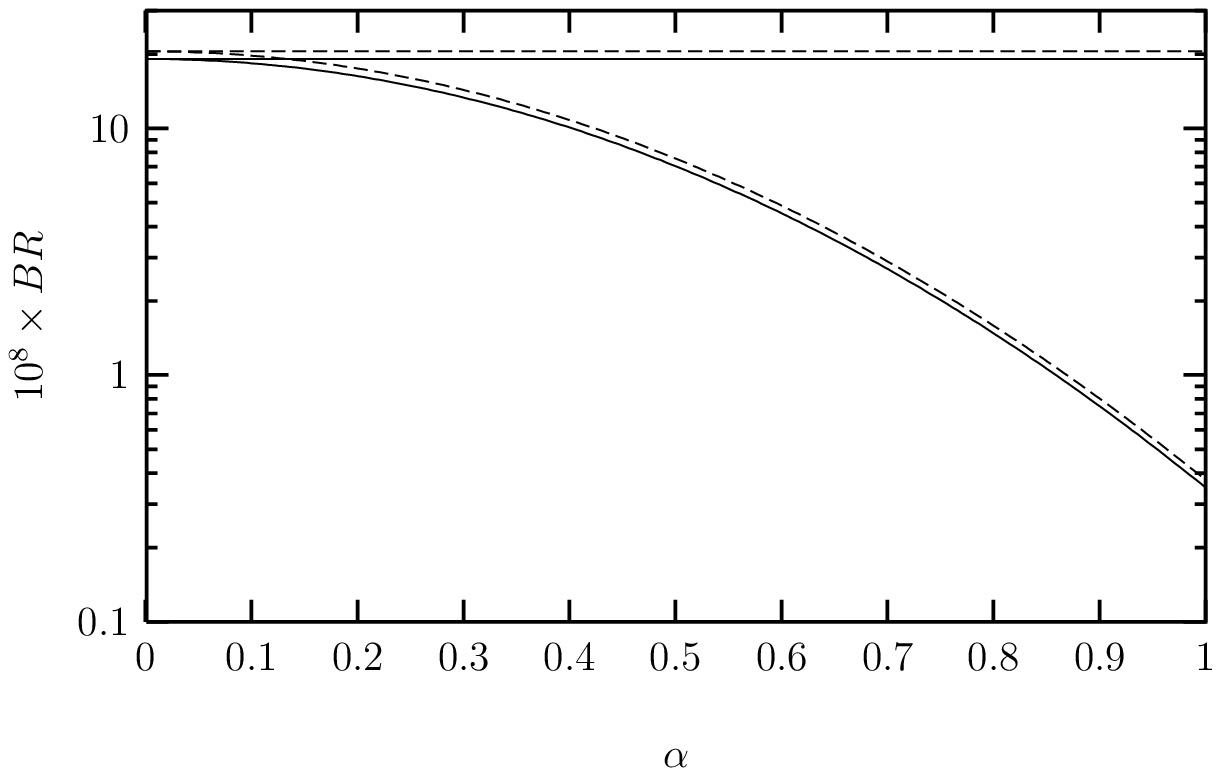} \vskip -3.0truein
\caption[]{BR($\tau\rightarrow \mu \gamma$) with respect to
$\alpha$ for $1/R= 1000\,GeV$ and $\bar{\xi}^{E}_{N,\tau \mu}
=10\, GeV$, $\bar{\xi}^{E}_{N,\tau \tau} =50\, GeV$. Here the
solid-dashed line (curve) represents the BR without-with lepton KK
mode contribution in the case that the new Higgs doublet is
located around the origin ($z_H=\alpha\,\sigma$) in the sixth
dimension.} \label{BRtaumugamalf}
\end{figure}

In Fig. \ref{BRtaumugamR} we present the BR($\tau\rightarrow \mu
\gamma$) with respect to the scale $1/R$. Here the solid-dashed
line (curve) represents the BR for $\bar{\xi}^{E}_{N,\tau \tau}
=50\, GeV$ and $\bar{\xi}^{E}_{N,\tau \mu} =1\, GeV$
-$\bar{\xi}^{E}_{N,\tau \mu} =10\, GeV$ without (with) lepton KK
mode contribution in the case that the new Higgs doublet is
located around the origin in the sixth dimension. The BR
($\tau\rightarrow \mu \gamma$) is at the order of magnitude of
$10^{-7}$ for the coupling $\bar{\xi}^{E}_{N,\tau \mu} =10\, GeV$.
With the addition of the internal lepton KK mode contributions it
enhances  more than $50\%$ for the compactification scale $1/R\sim
500\,GeV$. Fig. \ref{BRtaumugamalf} shows the scale $\alpha$
dependence of the BR($\tau\rightarrow \mu \gamma$) for $1/R=
1000\,GeV$ and $\bar{\xi}^{E}_{N,\tau \mu} =10\, GeV$,
$\bar{\xi}^{E}_{N,\tau \tau} =50\, GeV$. Here the solid-dashed
line (curve) represents the BR without-with lepton KK mode
contribution in the case that the new Higgs doublet is located
around the origin ($z_H=\alpha\,\sigma$) in the sixth dimension.
We observe that the BR  is suppressed more than one order of
magnitude for the case that the Gaussian profile is located one
$\sigma$ farther from the origin.
\\ \\
Now we would like to summarize our results:

\begin{itemize}
\item The inclusion of KK modes due to the fifth dimension results
in  enhancement in the electron (muon, tau) EDM at the order of
$25\%$ ($30\%$, $20\%$) for the compactification scale $1/R\sim
500\,GeV$.
\item Charged Lepton EDMs are strongly sensitive to the location
of new Higgs doublet Gaussian profile in the sixth dimension and
their magnitudes are suppressed almost one order of magnitude even
in the case that the Gaussian profile is located one $\sigma$
farther from the origin.
\item The BRs of LFV $l_1\rightarrow l_2 \gamma$ decays enhance
almost $50\%$ with the inclusion of KK modes due to the fifth
dimension for the compactification scale $1/R\sim 500\,GeV$. This
enhancement is the almost two orders of magnitude for the small
values of scale $1/R$.
\item The BRs of LFV $l_1\rightarrow l_2\gamma$ decays are
strongly sensitive to the location of new Higgs doublet Gaussian
profile in the sixth dimension and their magnitudes are suppressed
almost two orders of magnitude even in the case that the Gaussian
profile is located one $\sigma$ farther from the origin.
\end{itemize}

With the help of the forthcoming most accurate experimental
measurements, the valuable information can be obtained about the
existence of extra dimensions and the possibilities of Gaussian
profiles of the Higgs scalars.
\newpage
\end{document}